\documentclass{vgtc}                          

\ifpdf
  \pdfoutput=1\relax                   
  \usepackage{graphicx}                
  \DeclareGraphicsExtensions{.pdf,.png,.jpg,.jpeg} 
\else
  \usepackage{graphicx}                
  \DeclareGraphicsExtensions{.eps}     
\fi%

\graphicspath{{figures/}{pictures/}{images/}{./}} 

\usepackage{microtype}                 
\PassOptionsToPackage{warn}{textcomp}  
\usepackage{textcomp}                  
\usepackage{mathptmx}                  
\usepackage{times}                     
\usepackage{cite}                      
\usepackage{tabu}                      
\usepackage{booktabs}                  
\usepackage{pifont,subfig}
\usepackage{amsmath,amssymb,amsfonts}
\usepackage{algorithm,algpseudocode}
\usepackage{adjustbox}
\usepackage{multirow}

\onlineid{0}

\vgtccategory{Research}

\vgtcinsertpkg

\title{360TripleView: 360-Degree Video View Management System Driven by Convergence Value of Viewing Preferences}

\author{
Qian Zhou\thanks{e-mail: qianz@illinois.edu}\\ %
        \scriptsize University of Illinois Urbana-Champaign %
\and Michael Zink\thanks{e-mail: zink@ecs.umass.edu}\\ %
     \scriptsize University of Massachusetts Amherst %
\and Ramesh Sitaraman\thanks{e-mail: ramesh@cs.umass.edu}\\ %
     \scriptsize University of Massachusetts Amherst %
\and Klara Nahrstedt\thanks{e-mail: klara@illinois.edu}\\ %
     \scriptsize University of Illinois Urbana-Champaign %
     }

\abstract{360-degree video has become increasingly popular in content consumption. However, finding the viewing direction for important content within each frame poses a significant challenge. Existing approaches rely on either viewer input or algorithmic determination to select the viewing direction, but neither mode consistently outperforms the other in terms of content-importance. In this paper, we propose 360TripleView, the first view management system for 360-degree video that automatically infers and utilizes the better view mode for each frame, ultimately providing viewers with higher content-importance views. Through extensive experiments and a user study, we demonstrate that 360TripleView achieves over 90\% accuracy in inferring the better mode and significantly enhances content-importance compared to existing methods.}

\newcommand{\degree}{$^{\circ}$}
\newcommand{\vid}{360\degree ~video}
\newcommand{\eb}[1]{\emph{\textbf{#1}}}
\newcommand{\man}{\textsc{Manual}}
\newcommand{\weak}{\textsc{Auto}$^{\textsc{optional}}$}
\newcommand{\weakman}{\weak /\man}
\newcommand{\strong}{\textsc{Auto}$^{\textsc{enforced}}$}
\newcommand{\use}{$mode_{use}$}

\begin{document}


\maketitle
\section{Introduction}
Omnidirectional video (\vid) distinguishes itself from conventional 2D video by capturing a panoramic field of view. During playback, each 360\degree ~frame is projected onto a 2D view based on the viewer's viewing direction, resulting in an immersive and flexible viewing experience. \vid ~aligns naturally with the growing demand for virtual and mixed reality applications, which industry giants such as Apple, Google, Microsoft, Meta, and others are heavily investing in, making it an integral part of our content consumption.

However, a crucial challenge lies in finding a 2D view from each 360\degree ~frame that provides important content to the viewer. Note that: 1) each 360\degree ~frame offers multiple potential viewing directions, and thus multiple potential views; 2) the same view can hold varying degrees of content-importance to different viewers, owing to their diverse viewing preferences. As a result, unless all possible views are examined, and the viewer's preference is taken into account, the selected view displayed to them may lack the desired importance.

Various approaches have been developed to identify important 2D views, with most falling into two categories. Firstly, many works~\cite{park2020seaware,nasrabadi2020viewport,sun2020flocking,samiei2021improving,
rondon2021track,chao2021transformer,vats2022semantic,guimard2022deep,zhou2022360broadview} rely on manual viewer control to select views. In this approach, the viewer manually adjusts their viewing direction during playback, receiving the corresponding 2D view. By actively controlling the viewing direction, they can prioritize views that are highly important to them based on their individual viewing preference. However, due to the limited field of view of a human, the viewer can only observe one of the many possible views for each 360\degree ~frame. Consequently, while they can still discover important content within their field of view, they may overlook other views out of their sight that possess even higher content-importance.

Secondly, other works~\cite{su2016pano2vid,su2017making,hu2017deep,wang2020attention,lai2017semantic,yu2018deep,lee2018memory} rely on algorithmic approaches, often utilizing saliency detection~\cite{cheng2018cube,dahou2021atsal}. In this scenario, a video server performs saliency detection on each 360\degree ~frame to identify the most salient 2D view. By systematically examining all possible views for each frame, this algorithmic approach may discover views with higher content-importance compared to manually selected views. However, saliency detection algorithms do not consider the diverse viewing preferences of different viewers. As a result, the same view is recommended to all viewers, which can be of high importance to those who prefer to focus on salient objects but of lesser importance to viewers whose preferences do not align with saliency. While machine learning-based personalization could be considered to recommend different views to different viewers, it necessitates prior knowledge of each individual's \vid ~viewing preference, which is typically unavailable.

As a result, viewers face a dilemma: should they rely on their own instintcs to find views, which consistently offer high but not excellent content-importance, or watch algorithm-found views that fluctuate between excellent and low content-importance? Since neither view mode consistently outperforms the other in terms of content-importance, relying on a single mode throughout the video would result in limited content-importance. In this paper, we propose 360TripleView, the first intelligent \vid ~view management system that addresses this dilemma by dynamically inferring the better view mode for each 360\degree ~frame. 360TripleView offers three view modes, each serving a specific purpose:

\begin{itemize}
\item \textbf{\man.} Each viewer manually selects their views.
\item \textbf{\weak.} Algorithm-found views are provided, but viewers in \weak ~have the option to switch between \man ~and \weak.
\item \textbf{\strong.} Algorithm-found views are provided, and no manual intervention is permitted in \strong.
\end{itemize}

The key to enhancing the overall content-importance for viewers in 360TripleView lies in its \textbf{View Mode Decision-Maker}, which automatically determines whether to utilize \strong ~or \weakman ~for each 360\degree ~frame. It assesses whether the frame's algorithm-found views have higher content-importance than viewer-found views. If so, it employs \strong ~to ensure that everyone observes the algorithm-found views. Otherwise, it utilizes \weakman. In \weakman, 360TripleView suggests using \man ~since viewer-found views are deemed more important. However, \weak ~is also available as an option, allowing viewers to switch to it if they experience fatigue from prolonged manual control and prefer to enjoy algorithm-found views (they can switch back to \man ~at any time).

Inferring the better mode (in terms of higher content-importance) between \strong ~and \weakman ~is a problem that has not been previously studied. We tackle this challenge based on the following insights: 1) Viewers' viewing preferences exhibit convergence in some 360\degree ~frames while divergence in others. We quantify this degree of convergence using a novel metric we define as the \textbf{Convergence Value of Viewing Preferences (CVVP)}. 2) We find that CVVP is instrumental in inferring the better mode between \strong ~and \weakman. For instance, a low CVVP for a 360\degree ~frame indicates divergent opinions among viewers regarding the important view for that frame. In such cases, recommending a single view through \strong ~would only cater to a small fraction of viewers, limiting the overall satisfaction. Therefore, a low CVVP leads to the inference that \weakman ~is the better mode, allowing viewers to manually control their own viewing directions. 3) We demonstrate that an automatic CVVP estimator, which takes a 360\degree ~frame as input and provides an estimated CVVP, can be effectively trained, achieving high accuracy in CVVP estimation. Consequently, no viewer needs to provide their viewing preference when using 360TripleView.

Our contributions in this paper are as follows:
\begin{itemize}
\item In Section~\ref{sec:overview}, we introduce 360TripleView, the first view management system for \vid ~that enhances overall content-importance for viewers by automatically inferring and utilizing the better mode between \strong ~and \weakman ~for each 360\degree ~frame.

\item In Section~\ref{sec:cvvp}, we define the CVVP metric and propose a deep learning-based solution to estimate the CVVP for each frame automatically. During the offline model training stage, a few viewers' labeled viewing preferences on some videos are required to generate the ground truth CVVP. In the online stage, the model takes frames from new videos (not included in the training set) as input and returns the estimated CVVP, and no viewer needs to provide their viewing preference.

\item Our experiments (Section~\ref{sec:eval}) and user study (Section~\ref{sec:user}) demonstrate that 360TripleView achieves an accuracy above 90\% in inferring the better mode and delivers views of significantly higher content-importance compared to existing approaches.
\end{itemize}

We provide background information and discuss related work in Section~\ref{sec:related}, outline potential directions for future research in Section~\ref{sec:discussion}, and conclude our work in Section~\ref{sec:conclusion}.

\section{Background and Related Work} \label{sec:related}
\begin{figure}[h] 
\centering
\includegraphics[width=0.9\linewidth]{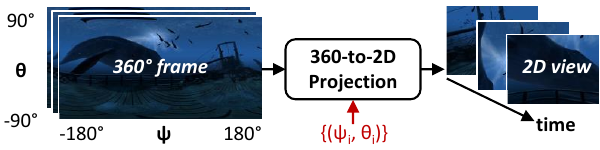}
\caption{Projection of 360\degree ~frames onto 2D views based on a sequence of viewing directions.}
\label{fig:background}
\end{figure}

\textbf{360\degree ~Video Viewing.} The nature of \vid ~is depicted in Fig.~\ref{fig:background}, where it is an omnidirectional recording with a much wider field of view (\textbf{FoV}) than that of human eyes (horizontally $<$ 120\degree). To make \vid ~viewing intuitive, a sequence of viewing directions $\{ (\psi_i, \theta_i) \}$ is provided, where the yaw angle $\psi_i \in [-180^{\circ}, 180^{\circ}]$, the pitch angle $\theta_i \in [-90^{\circ}, 90^{\circ}]$, and $i$ represents the frame ID. These viewing directions allow each 360\degree ~frame to be projected onto a 2D view for viewing. In this paper, the terms ``find a 2D view" and ``find the viewing direction $(\psi, \theta)$" are used interchangeably since determining the viewing direction of a 360\degree ~frame enables the identification of the corresponding 2D view.

\textbf{Content-Importance.} The importance of content within a 2D view varies according to the diverse viewing preferences of viewers. In this paper, we focus on developing a view management system that automatically infers the better view mode to utilize, making viewers obtain views with higher content-importance overall.

Existing view modes can be categorized as follows:
\subsection{\man ~Mode}
In \man ~mode, the viewer has complete control over their viewing direction $(\psi, \theta)$. Using a client-device (e.g., a mouse/keyboard or headset), the viewer manually adjusts their viewing direction throughout the video. The client-device continuously sends the updated $(\psi, \theta)$ to the video server, which keeps returning the corresponding 2D view to the viewer.

\textbf{Pros \& Cons.} Many existing works~\cite{qian2018flare,park2019navigation,
park2020seaware,nasrabadi2020viewport,sun2020flocking,samiei2021improving,
rondon2021track,chao2021transformer,vats2022semantic,guimard2022deep,
zhou2022360broadview} consider \man ~to be satisfactory for viewers. This is because manual control allows viewers to obtain views that align with their own preferences and are therefore important to them. However, due to the viewer's limited FoV, they can only see a small portion of the entire 360\degree ~frame at a given time. Consequently, they may perceive the view they are watching as the most important, while a more significant view exists outside of their sight, which they would have turned to if they had been aware of it. As a result, \eb{\man ~consistently delivers views of high content-importance to the viewer but not of excellent importance}.

To address the drawback of the limited FoV, Lin et al.~\cite{lin2017tell} introduce graphical indicators within the viewer's FoV to indicate targets outside their sight. Another approach, Outside-In~\cite{lin2017outside}, proposes advanced thumbnails for out-of-sight targets. These thumbnails are perspective projections that correspond to the positions of the targets, rather than simple rectangles. \eb{However, these solutions require the viewer to manually track multiple view candidates, which can be distracting or overwhelming}.

\subsection{\strong ~Mode}
On the other hand, \strong ~mode offers no control to the viewer. In this mode, the server performs saliency detection~\cite{cheng2018cube,dahou2021atsal} to determine the viewing direction $(\psi, \theta)$ that results in the most salient 2D view, which is then delivered to the viewer. In \strong, the viewer can only watch the auto-generated 2D video and is unable to switch to another mode.

The pioneering work Pano2Vid~\cite{su2016pano2vid} converts a \vid ~into a 2D video that resembles those captured by human videographers. Initially, it employs a fixed FoV of 65.5\degree, which is made variable in the authors' subsequent work~\cite{su2017making}. Deep360Pilot~\cite{hu2017deep} utilizes supervised learning, where the authors manually label the most salient object frame by frame and train an RNN to recommend the corresponding $(\psi, \theta)$. Wang et al.~\cite{wang2020attention} employ reinforcement learning using ground truth data from the Pano2Vid and Deep360Pilot datasets, combined with saliency detection. Lai et al.~\cite{lai2017semantic} incorporate saliency detection and semantic segmentation. They also propose saliency-aware temporal summarization, which increases the playback speed of frames with lower saliency scores.

\textbf{Pros \& Cons.} Since the algorithm considers the entire 360\degree ~frame and explores all possible 2D views, it may discover a view of higher content-importance compared to what a viewer might find in \man ~mode. However, due to the diverse viewing preferences among viewers, an algorithm-found view may be of excellent importance to some viewers but have little or no importance to others. Personalization through machine learning, which recommends different views to different viewers based on their preferences, is not feasible at this stage since obtaining every individual's \vid ~viewing preference is impractical. Therefore, in this paper, we assume that \eb{\strong ~recommends the same 2D view to all viewers who watch the same 360\degree ~frame, without personalization}.

As a result, viewers using \strong ~mode risk receiving a view that may be more or less important compared to what they would obtain in \man ~mode. In other words, \eb{neither \strong ~nor \man ~is consistently superior to the other in terms of content-importance}.

\subsection{\weakman ~Mode}
In \weakman ~mode, the viewer has the option to manually switch between algorithm-found views (\weak) and viewer-found views (\man) whenever desired. The algorithm-found views in \weak ~are the same as those in \strong. The key distinction is that the viewer can interrupt \weak ~and switch to \man, but such interruptions are not possible in \strong. By default, \weak ~is used, and if the viewer does not perform any operations for a while in \man, \weak ~will resume.

Kang et al.~\cite{kang2019interactive} employ saliency detection to recommend a sequence of $(\psi, \theta)$ with the highest cumulative saliency score. If the viewer chooses a lower-saliency view in \man, \weak ~will resume more slowly. Similarly, Cha et al.~\cite{cha2020enhanced} utilize saliency detection and face detection to recommend a $(\psi, \theta)$ sequence. In their approach, if the viewer changes their viewing direction more drastically in \man, \weak ~will resume more slowly. Transitioning360~\cite{wang2020transitioning360} relies on saliency detection and semantic segmentation to generate multiple $(\psi, \theta)$ sequences for \weak. The corresponding thumbnails of these sequences appear simultaneously in the viewer's FoV, and the viewer can click on a thumbnail to switch to the corresponding view.

These works aim to enhance content-importance by providing viewers with two mode options, allowing them to choose the better one. However, \eb{they overlook the fact that humans are incapable of accurately and instantly determining the better mode for each frame without being distracted from enjoying the video content}.

\section{Overview of 360TripleView} \label{sec:overview}
360TripleView is a groundbreaking view management system for \vid ~viewing. It has three view modes (\man, \weak, \strong), \eb{utilizing one mode at a time}. This innovative viewing system automatically infers and utilizes the better mode between \strong ~and \weakman, enabling viewers to experience views of enhanced content-importance. In \weakman, each viewer has the additional freedom to choose between \man ~and \weak ~without affecting the view modes of other viewers.

\subsection{Triple View Modes}
360TripleView offers three view modes:

\begin{itemize}
\item \textbf{\man.} Each viewer manually selects their views.
\item \textbf{\weak.} Algorithm-found views are provided, but viewers in \weak ~have the option to switch between \man ~and \weak.
\item \textbf{\strong.} Algorithm-found views are provided, and no manual intervention is permitted in \strong.
\end{itemize}

\begin{figure}[h] \centering
\includegraphics[width=0.9\linewidth]{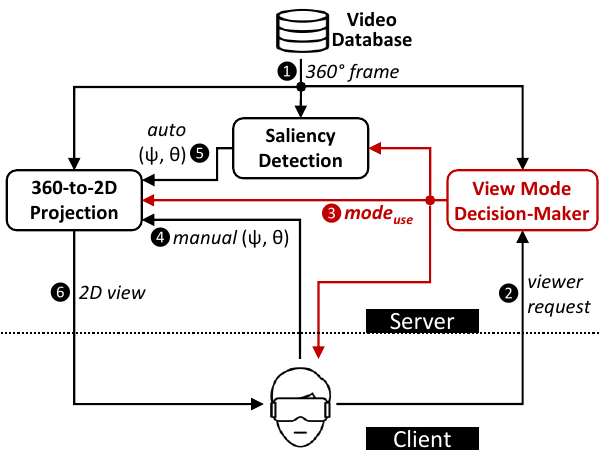}
\caption{360TripleView system architecture.}
\label{fig:overview1}
\end{figure}

As illustrated in Fig.~\ref{fig:overview1}, 360TripleView involves the server processing and transmitting video content, while viewers access the content through their client-devices (e.g., headsets). The \textbf{View Mode Decision-Maker} on the server receives two inputs: \ding{202} a 360\degree ~frame from the Video Database, and \ding{203} the viewer's request to change their view mode. The decision-maker determines \ding{204} the view mode to use for each frame and sends the decision (denoted as \use ~$\in \{$ \man, \weak, \strong $\}$) to the viewer, the saliency detection unit, and the 360-to-2D projection unit. If \use ~is \man, the viewer continues to provide \ding{205} their manually controlled viewing direction $(\psi, \theta)$ to the projection unit; if \use ~is \weak ~or \strong, the saliency detection unit processes the video frame and \ding{206} automatically recommends a $(\psi, \theta)$ to the projection unit. It is important to note that 360TripleView can utilize any existing saliency detection approach (e.g., we use ATSal~\cite{dahou2021atsal}, but other solutions also work). Finally, the projection unit receives the video frame and the $(\psi, \theta)$ (either \ding{205} or \ding{206}, depending on \use), and delivers \ding{207} the corresponding 2D view to the viewer.

\subsection{View Mode Decision-Maker}
\begin{figure}[h] \centering
\includegraphics[width=0.9\linewidth]{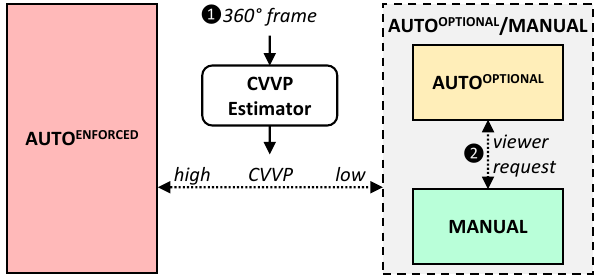}
\caption{360TripleView view mode decision-maker.}
\label{fig:overview2}
\end{figure}

The cornerstone of 360TripleView is its View Mode Decision-Maker (Fig.~\ref{fig:overview2}). This component determines, for each 360\degree ~frame, which of the three view modes (\man, \weak, \strong) to utilize (i.e., \use).

As depicted in Fig.~\ref{fig:overview2}, the View Mode Decision-Maker operates as a state machine with three states: \man, \weak, \strong. The \textbf{CVVP Estimator} is responsible for determining whether to employ \strong ~or \weakman. It takes \ding{202} a 360\degree ~frame as input and estimates the \textbf{Convergence Value of Viewer Preferences (CVVP)}. The CVVP is a metric that indicates the better mode, either \strong ~or \weakman, in terms of higher content-importance. We will provide further details on this metric in Section~\ref{sec:cvvp}. If the CVVP exceeds a certain threshold (e.g., $> 60\%$, which is configurable), it suggests that algorithm-found views may offer higher overall content-importance compared to viewer-found views. Therefore, in such cases, \use ~will be set to \strong ~to ensure that algorithm-found views are not missed by any viewer. Conversely, a small CVVP implies that viewer-found views possess higher content-importance. Consequently, \use ~will be set to \weakman, allowing each viewer to manually send \ding{203} a viewer request (e.g., through mouse clicks or dragging) to the View Mode Decision-Maker (Fig.~\ref{fig:overview2}) and switch the \use ~between \man ~and \weak ~(without affecting other viewers' \use). When viewers select \man, they watch the personally selected views.

\textbf{Why is \weak ~Useful?} One might question the necessity of \weak: why not utilize only two modes (\strong ~and \man) and let \use ~simply be \strong ~if algorithm-found views are deemed more important, and \man ~otherwise? The reason is that viewers in \man ~are required to constantly engage in manual control to change their viewing directions. Without continuous input, their viewing directions will remain static. We introduce \weakman ~to substitute for \man ~and alleviate this burden, allowing viewers the option to watch algorithm-found views (\weak) if desired, especially when they feel fatigued after prolonged use of \man. Moreover, visual cues are provided to remind viewers that algorithm-found views may currently have lower content-importance than viewer-found views, suggesting that they use \man ~to obtain views of higher content-importance.

\section{View Mode Decision-Making Driven by CVVP} \label{sec:cvvp}
The key to enhancing the overall content-importance for viewers in 360TripleView lies in its View Mode Decision-Maker, which autonomously infers whether \strong ~or \weakman ~will yield higher content-importance. How is such inference achieved? In this section, we introduce a novel metric known as the Convergence Value of Viewer Preferences (CVVP). We explain how this metric is utilized to infer the better mode and present a deep learning solution to automatically estimate the CVVP for each 360\degree ~frame, automating the entire inference process.

\subsection{Definition of CVVP} \label{sec:cvvp_1}
\begin{figure}[h] \centering
\includegraphics[width=\linewidth]{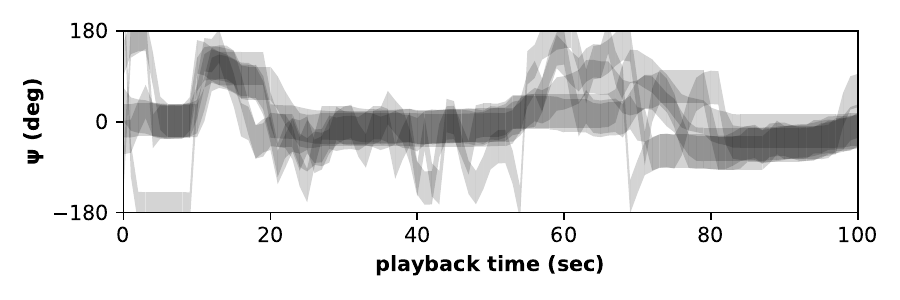}
\caption{Variation of viewers' labeled $\psi$ (yaw), showing convergence and divergence. The phenomenon also applies to $\theta$ (pitch).}
\label{fig:design1}
\end{figure}

Through experiments, we have made a key observation: when multiple viewers are presented with a complete view of a 360\degree ~frame, and each viewer is asked to identify the viewing direction that holds their highest content-importance. Their preferences---indicated by their labeled directions---diverge in some frames while converge in others. Figure~\ref{fig:design1} illustrates the variation of $\psi$ (yaw) labeled by 6 viewers over time in a video from the Pano2Vid~\cite{su2016pano2vid} dataset (\texttt{www.youtube.com/watch?v=i9SiIyCyRM0}): their preferences diverge from second 57 to 80 and converge well at other times. Divergence often occurs when a frame contains \emph{zero or multiple} significant regions, leading to different choices based on individual preferences. Conversely, convergence occurs when a frame contains \emph{one} dominant important region (e.g., a host speaking to the camera in a tour video, a player attempting a shot in a sports video, a whale approaching in an underwater video), which is favored by most viewers. Consequently, viewers' viewing preferences exhibit a dynamic degree of convergence that varies across video frames.

It is noteworthy that some works~\cite{almquist2018prefetch,sitzmann2018saliency,rossi2019spherical,
rossi2020understanding,nasrabadi2020viewport,rondon2021track,
hirway2022spatial} have examined viewers' behaviors and patterns during \vid ~viewing, but they solely investigate \man ~mode. These studies observe dynamic convergence in the viewports of viewers in \man, where viewers have a limited field of view. In contrast, we reveal that the viewing preferences of viewers, labeled by viewers when provided with a complete 360\degree ~view, exhibit dynamic convergence.

To quantify the degree of convergence for each 360\degree ~frame, we introduce a novel metric called the Convergence Value of Viewer Preferences (CVVP). Computing the \textbf{ground truth CVVP} requires knowledge of the most important labeled viewing directions from $N$ viewers, denoted as $(\psi_j, \theta_j)$, where the viewer ID is represented by $j = 1, 2, \ldots, N$. The \textbf{content-importance} of a viewing direction $(\psi, \theta)$ to viewer $j$ is defined as follows:
\begin{equation}
importance_j(\psi, \theta) = 
\begin{cases}
1 \quad \text{if} ~\text{gcd}((\psi, \theta), (\psi_j, \theta_j)) < TH_{dist} \\
0 \quad \text{otherwise}
\end{cases}
\end{equation}
where gcd$()$ calculates the great-circle distance between two viewing directions (each direction corresponds to a point on the unit sphere), and $TH_{dist}$ is the distance threshold. Specifically, $(\psi, \theta)$ is considered important to viewer $j$ if it is sufficiently close to the viewer's labeled direction. Given that the human field of view is $< 120$\degree, we consider two directions to be close enough if they are less than 30\degree ~apart. Thus, we set $TH_{dist}$ to 30\degree.

The \textbf{overall content-importance} of $(\psi, \theta)$ is defined as the average of the content-importance values across all $N$ viewers:
\begin{equation} \label{equ:2}
importance(\psi, \theta) = \frac{1}{N}\sum_{j=1}^{N} importance_j(\psi, \theta)
\end{equation}

Finally, the \textbf{CVVP} of the 360\degree ~frame is defined as the maximum $importance(\psi, \theta)$ among all $\psi \in [-180^{\circ}, 180^{\circ}]$, $\theta \in [-90^{\circ}, 90^{\circ}]$:
\begin{equation}
\text{CVVP} = \underset{\psi, \theta}{\max} ~importance(\psi, \theta)
\end{equation}

\begin{figure}[h] \centering
\subfloat[CVVP $=\frac{2}{6} \approx 0.33$]{\includegraphics[width=0.495\linewidth]{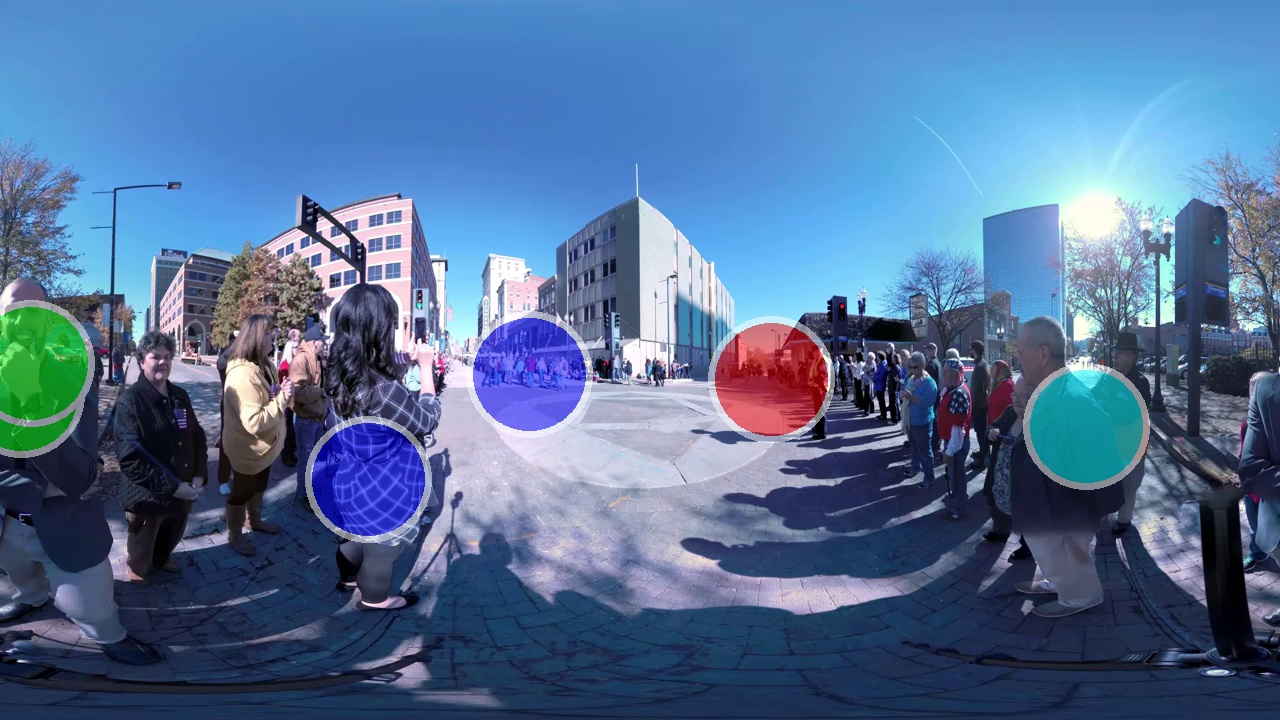}} \hfill
\subfloat[CVVP $=\frac{6}{6}=1$]{\includegraphics[width=0.495\linewidth]{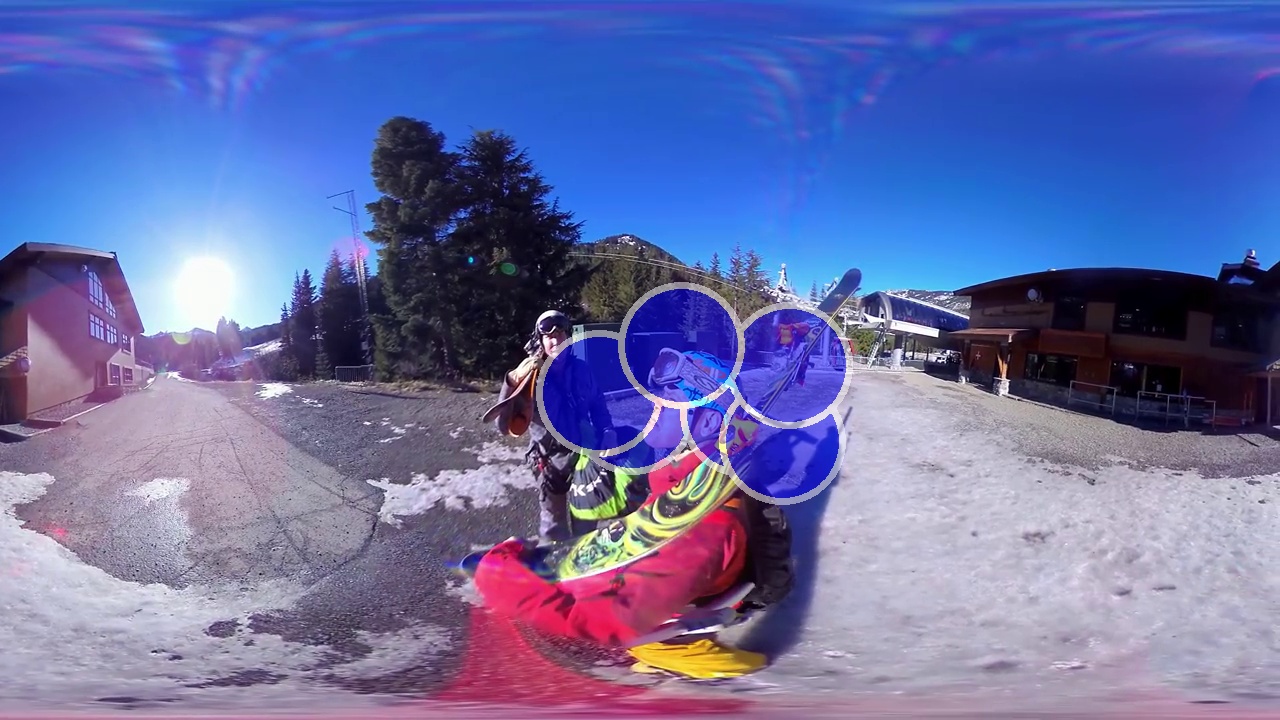}} \hfill
\caption{Examples of ground truth CVVP. Each circle represents a viewer-labeled most important viewing direction. Directions close to each other are color-coded identically.}
\label{fig:demo}
\end{figure}

Note that we require $N$ viewers to label their viewing preferences for some frames in order to generate the ground truth CVVP. This ground truth is only used to train and test the CVVP Estimator offline (Section~\ref{sec:cvvp_3}). When the estimator runs online for new videos, neither prior nor current viewers need to provide any labels or viewing preferences (details can be found in Section~\ref{sec:cvvp_3}). Figure~\ref{fig:demo} presents two examples of frames, their labels and ground truth CVVP values.

\subsection{Using CVVP to Infer the Better Mode} \label{sec:cvvp_2}
The View Mode Decision-Maker utilizes CVVP to infer the better mode (in terms of higher overall content-importance), between \strong ~and \weakman. The inferred better mode is then selected as the mode to use (\use):
\begin{equation}
\text{\use} = 
\begin{cases}
\text{\weakman} \quad \text{if CVVP} < TH_{CVVP} \\
\text{\strong} \quad \quad \quad \quad \quad \text{otherwise}
\end{cases}
\end{equation}
where $TH_{CVVP}$ represents a configurable CVVP threshold.

Why is CVVP an effective indicator for inferring the better mode? Note that CVVP $\in (0, 1]$ increases with the convergence degree of viewing preferences:

\begin{enumerate}
\item In the case of the lowest convergence, where $(\psi_j, \theta_j)$ are scattered, any $(\psi, \theta)$ is close to at most one $(\psi_j, \theta_j)$, i.e.,

\begin{equation*}
\forall (\psi, \theta), ~importance(\psi, \theta) \le 1, \quad \text{thus CVVP} ~= \frac{1}{N}
\end{equation*}

\item In the case of the highest convergence, where all $N$ viewers' $(\psi_j, \theta_j)$ are closely clustered, there exists a $(\psi, \theta)$ that is close to all of them, i.e.,

\begin{equation*}
\exists (\psi, \theta), ~importance(\psi, \theta) = N, \quad \text{thus CVVP} ~=1
\end{equation*}

\item In general, if a 360\degree ~frame has a CVVP of $\eta \%$, it implies that at most $\eta \%$ of the viewers will have their preferred view when one $(\psi, \theta)$ is viewed by all viewers. Notably, \strong ~recommends one $(\psi, \theta)$ to all viewers without personalization (Section~\ref{sec:related}). Thus, CVVP serves as an upper bound for the actual overall content-importance achieved by \strong ~for that specific frame.
\end{enumerate}

Based on these observations, the inferences below can be made:
\begin{itemize}
\item If CVVP $< TH_{CVVP}$ (e.g., 60\%), the actual overall content-importance achieved by \strong ~must be $< TH_{CVVP}$. Thus, \weakman ~is inferred as the better mode and becomes \use, allowing viewers to use \man.

\item If CVVP $\geq TH_{CVVP}$, the actual overall content-importance achieved by \strong ~can be $\geq TH_{CVVP}$. Thus, \strong ~is inferred as the better mode and becomes \use, ensuring that algorithm-found views are watched.
\end{itemize}

\subsection{Automatic CVVP Estimator} \label{sec:cvvp_3}
To compute the ground truth CVVP (Section~\ref{sec:cvvp_1}) of a 360\degree ~frame, we need to know the important viewing directions labeled by multiple viewers for that frame. However, obtaining this information from viewers during the use of 360TripleView is impractical. Therefore, we introduce a deep learning solution: in the model training stage (offline), we request a few viewers to label some frames of some videos, enabling us to compute the ground truth CVVP; during the online usage, the model processes frames from new videos (not used in training) and provides the estimated CVVP without requiring any viewers to provide their viewing preferences.

\subsubsection{Deep Learning-Based Regression} \label{sec:cvvp_31}
We have developed a deep learning-based regression model that takes a 360\degree ~frame (represented as $I$) as input and predicts its CVVP $\in \mathbb{R}$ and $\in (0, 1]$. The model leverages ResNet101~\cite{he2016deep} for image feature extraction. Since ResNet is pretrained on 2D images and does not handle equirectangular 360\degree ~frames, which are significantly distorted in the polar regions, we first convert $I$ to a cubemap comprising six 2D views denoted as $\{I^{x}\}$, where $x = \text{front, back, left, right, up, down}$. Each view is passed through the feature extractor to obtain its visual features $V_{I^x} \in \mathbb{R}^{2048}$. These features are then concatenated to form $V_I \in \mathbb{R}^{12288}$.

\begin{figure}[h] \centering
\includegraphics[width=\linewidth]{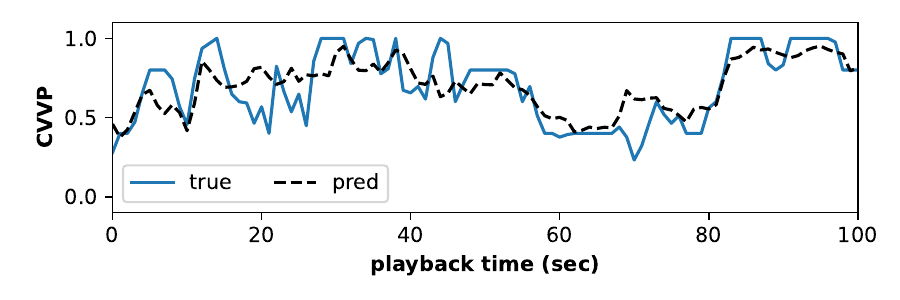}
\caption{CVVP predicted by the neural network.}
\label{fig:design2}
\end{figure}

$V_I$ is fed through two fully connected layers with output sizes of 2048 and 1 (representing the predicted CVVP), respectively. ReLU activation is applied to each layer. Mean Absolute Error (L1) is used as the loss function instead of Mean Squared Error (L2) because MAE is more robust against outliers. Fig.~\ref{fig:design2} displays the predicted CVVP and the ground truth for the video in Fig.~\ref{fig:design1}. It can be observed that the predicted and ground truth values are relatively low from second 57 to 80, consistent with the divergence depicted in Fig.~\ref{fig:design1}.

\subsubsection{Binarization and Stabilization} \label{sec:cvvp_32}
\begin{figure}[h] \centering
\includegraphics[width=\linewidth]{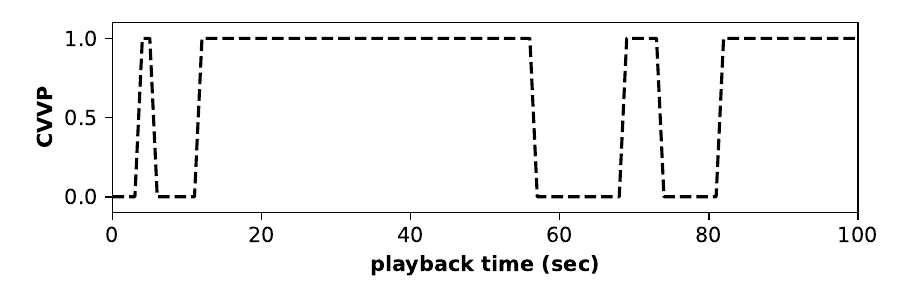}
\caption{Predicted CVVP sequence without stabilization.}
\label{fig:design3}
\end{figure}

The regression model outputs a sequence of CVVP values $\{CVVP_i \}$, where $i$ represents the frame ID. Note that if \use ~changes with CVVP frame by frame, viewers will be disturbed. To stabilize it, we compute the average CVVP per second, getting $\{CVVP_t \}$ where $t=1,2,\ldots,T$ second, $CVVP_t \in \mathbb{R}$ and $\in (0, 1]$. However, we notice that $CVVP_t$ still fluctuates often. If we simply binarize the $\{CVVP_t \}$ in Fig.~\ref{fig:design2} using threshold $TH_{CVVP}$ (e.g., 0.6, Section~\ref{sec:cvvp_2}), making a CVVP above $TH_{CVVP}$ be 1 (indicating the use of \strong) and otherwise be 0 (indicating the use of \weak/\man), we will get the sequence shown in Fig.~\ref{fig:design3},  with overly short view modes and frequent view mode switching. To address this issue, we devise an advanced stabilization module.

The stabilization module takes the sequence $\{CVVP_t \}$ as input, along with two parameters set by the system administrator: 1) $TH_{CVVP}$, the threshold above which a CVVP is considered large enough to use \strong, and 2) $t_{min}$, the minimum duration (e.g., 20 seconds) for which the view mode should remain unchanged before it can change again, ensuring stability.

First, we normalize $\{CVVP_t \}$ to $\{CVVP_t' \}$ such that $CVVP_t = TH_{CVVP}$ is mapped to 0.5. Then, we binarize $\{CVVP_t' \}$ to $\{\overline{CVVP}_t \}$, where $t=1,2,\ldots,T$ represents the second and $\overline{CVVP}_t \in \{0, 1\}$. The goal is to minimize the difference between $\{\overline{CVVP}_t \}$ and $\{CVVP_t' \}$ while ensuring that each time $\overline{CVVP}_t$ changes (from 0 to 1 or 1 to 0), the new value persists for at least $t_{min}$ seconds.

A brute-force search would require $2^T$ attempts. However, note that $\{\overline{CVVP}_t \}$ consists of disjoint subsequences, with each being either all 0 or all 1, and at least $t_{min}$ seconds long. So, there are at most $\left\lfloor \frac{T}{t_{min}} \right\rfloor$ subsequences. We can formulate the problem as follows:

\begin{equation}
\begin{aligned}
\underset{\{\overline{CVVP}_t \}_{v,[t_1,t_2,\ldots,t_m]}}{\arg\min} \quad & \text{MSE}(\{\overline{CVVP}_t \}_{v,[t_1,t_2,\ldots,t_m]}, \{CVVP_t' \}) \\
\textrm{s.t.} \quad & v \in \{0,1\} \\
\quad & \sum_{i=1}^m {t_i} = T \quad 1 \le m \le \left\lfloor \frac{T}{t_{min}} \right\rfloor, t_i \ge t_{min} \\
\end{aligned}
\end{equation}

where $v$ represents the initial value of $\{\overline{CVVP}_t \}$ (0 or 1), and $[t_1,t_2,\ldots,t_m]$ indicates that $\{\overline{CVVP}_t \}$ consists of $m$ disjoint subsequences, with $t_i$ being the length of the $i$th subsequence. Thus, $\{\overline{CVVP}_t \}_{v,[t_1,t_2,\ldots,t_m]}$ represents the sequence that starts with the value $v$, lasts for $t_1$ seconds, toggles the value (0 to 1 or 1 to 0), lasts for $t_2$ seconds, and so on. The second constraint ensures that all $t_i$ add up to $T$ (the total length of the sequence) and that each subsequence is at least $t_{min}$ seconds long. The optimal solution, which minimizes the mean squared error (MSE) to $\{CVVP_t' \}$, is the resulting sequence $\{\overline{CVVP}_t \}$.

The complexity of finding the optimal sequence depends on the number of candidates to test. We can consider this as a ``combinations of repetition" problem. Cutting a sequence of $T$ seconds into $m$ subsequences is equivalent to inserting $(m-1)$ separators into the sequence. This is equivalent to selecting $(m-1)$ positions out of $(T+m-1)$ positions to place the separators, resulting in ${T+m-1 \choose m-1}$ possible combinations. However, this equation allows adjacent separators, which can result in subsequences of 0 seconds. To ensure that each subsequence is at least $t_{min}$ seconds long, we need to consider ${T-m \cdot t_{min}+m-1 \choose m-1}$ combinations. Additionally, $m$ can be any integer between 1 and $\left\lfloor \frac{T}{t_{min}} \right\rfloor$, and the initial value $v$ has 2 options (0 or 1). Considering all these factors, the total number of sequence candidates to be tested to find the optimal solution is:

\begin{equation}
2\sum_{m=1}^{\left\lfloor \frac{T}{t_{min}} \right\rfloor} {T-m \cdot t_{min}+m-1 \choose m-1}
\end{equation}

\begin{figure}[h] \centering
\includegraphics[width=\linewidth]{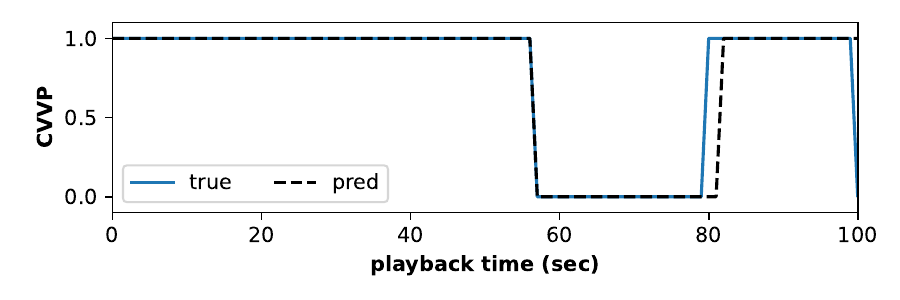}
\caption{CVVP sequence with stabilization.}
\label{fig:design4}
\end{figure}

The number of candidates decreases as $T$ decreases or $t_{min}$ increases. In our case, with $T = 120$ seconds and $t_{min} = 20$ seconds, the video is divided into 120-second clips, and each clip is further divided into view mode segments that last for at least 20 seconds. This results in 49,882 candidates, and the optimal solution can be found efficiently. Choosing a smaller $t_{min}$, such as 10 seconds, can lead to more frequent mode switching and discomfort for viewers. On the other hand, selecting a larger $t_{min}$, such as 30 seconds, may result in a significant error between $\{\overline{CVVP}_t \}$ and $\{CVVP_t' \}$. The stabilized result of the sequence in Fig.~\ref{fig:design3} is illustrated in Fig.~\ref{fig:design4}.

\section{Experiments} \label{sec:eval}
\subsection{Implementation and Experimental Settings}
\subsubsection{Dataset}
To train and test the deep learning model, we acquire \vid s with ground truth CVVP. We utilize the Pano2Vid dataset~\cite{su2016pano2vid}, which has 15 videos of diverse content, such as tours, sports, and parades. Each frame of each video in Pano2Vid has 6 most important $(\psi, \theta)$ labeled by different participants (see examples in Fig.~\ref{fig:demo}). We convert these labels to ground truth CVVP per frame, following the definition of CVVP (Section~\ref{sec:cvvp_1}). It is worth noting that \emph{each participant is given the entire view of a 360\degree ~frame and asked to check every viewing direction and label the highest content-importance view according to their viewing preference, frame by frame.}

Pano2Vid is the only dataset that meets our criterion for ground truth CVVP generation. Most other \vid ~datasets~\cite{corbillon2017360,wu2017dataset,lo2017360,rai2017dataset,fremerey2018avtrack360, david2018dataset,nasrabadi2019taxonomy,nguyen2019saliency,chakareski2021full,guimard2022pem360,jin2022you} are collected by having participants watch videos with headsets in \man ~mode while recording their viewing directions in real time. However, \man ~cannot guarantee that the most important $(\psi, \theta)$ is found due to the limited field of view. Therefore, these datasets are unsuitable for generating ground truth CVVP. Another dataset, Deep360Pilot~\cite{hu2017deep}, provides participants with the entire 360\degree ~view for labeling. However, it only offers one important $(\psi, \theta)$ label per frame, while CVVP ground truth generation requires multiple viewers' $(\psi, \theta)$ as input.


\subsubsection{Deep Learning} \label{sec:tuning}
For feature extraction in our CVVP regression model, we use ResNet101~\cite{he2016deep}. We have experimented with other models, such as VGG19~\cite{simonyan2014very} and Inception-v3~\cite{szegedy2016rethinking}, but found no significant impact on the accuracy of CVVP estimation. We employ stochastic gradient descent with a learning rate of 0.001 and momentum of 0.9 for training, running for 100 epochs.

We conduct three training and testing schemes:

\begin{itemize}
\item \textbf{No Tuning (leave-one-out):} Each video is tested using the model trained on the other videos. This is a commonly used cross-validation scheme. However, due to the limited size and diverse content of the Pano2Vid dataset, a video for testing often looks dissimilar to the other 14 videos for training. As a result, this scheme can yield poor testing results, which would not occur if the dataset were larger. Therefore, the performance of our machine learning solution may be underestimated. So, we also provide two fine-tuning schemes below.
\item \textbf{1-sec Tuning:} For each video, we randomly select 1 second (30 frames) of its content and use the corresponding ground truth CVVP to fine-tune the model trained on the other 14 videos. We then test this video using the fine-tuned model. We argue that if the training set were larger, no tuning would be necessary. Furthermore, even if a new video requires a tuned model for higher accuracy, acquiring 1-second ground truth is relatively easy, making our solution practical.
\item \textbf{3-sec Tuning:} This scheme is similar to 1-sec Tuning, but we use 3-second ground truth CVVP for fine-tuning.
\end{itemize}

\subsubsection{Evaluation Metrics}
We employ three metrics for evaluation:

\begin{itemize}
\item \textbf{Error of Estimated CVVP:} This metric represents the difference between the predicted CVVP per frame ($CVVP_i$, where $i$ is the frame ID) and the ground truth CVVP.
\item \textbf{Accuracy of Better Mode Inference:} A true positive (TP) occurs when both the predicted and ground truth CVVPs are 1 (indicating the use of \strong). A true negative (TN) occurs when both CVVPs are 0 (indicating the use of \weakman). As usual, $accuracy = \frac{TP+TN}{\text{total}}$.
\item \textbf{Overall Content-Importance:} It measures the actual overall content-importance, as defined in Equation~\ref{equ:2} in Section~\ref{sec:cvvp_1}.
\end{itemize}

\subsubsection{Baselines}

360TripleView automatically estimates $\{\overline{CVVP}_t\}$, determines \use ~as \strong ~when $\overline{CVVP}_t = 1$, and as \weakman ~when $\overline{CVVP}_t = 0$. We compare it with two baseline methods that determine \use:

\begin{itemize}
\item \textbf{\strong ~ONLY:} \use ~is always \strong.
\item \textbf{\weakman ~ONLY:} \use ~randomly switches between \weak ~and \man.
\end{itemize}

Note that 360TripleView innovates in \use ~determination (i.e., deciding which mode to use), not in saliency detection. When \use ~becomes \strong ~or \weak, the saliency detection unit (Fig.~\ref{fig:overview1}) executes an existing saliency detection algorithm, and its performance impacts the resulting overall content-importance. We test the following saliency detection approaches:

\begin{itemize}
\item \textbf{CubePad:} CubePadding~\cite{cheng2018cube} is a seminal and well-known saliency detection approach for \vid.
\item \textbf{ATSal:} ATSal~\cite{dahou2021atsal} is one of the most recent state-of-the-art saliency detection methods for \vid.
\item \textbf{Pano2Vid:} The $(\psi, \theta)$ in Pano2Vid~\cite{su2016pano2vid} are manually labeled, not algorithmically derived like CubePad or ATSal. We include it here because it represents the ``ceiling" of $(\psi, \theta)$ recommendation, which may be approached by future algorithms.
\end{itemize}

\subsection{Error of Estimated CVVP}
\begin{figure}[h] \centering
\subfloat[Error of estimated CVVP.]{\includegraphics[width=0.49\linewidth]{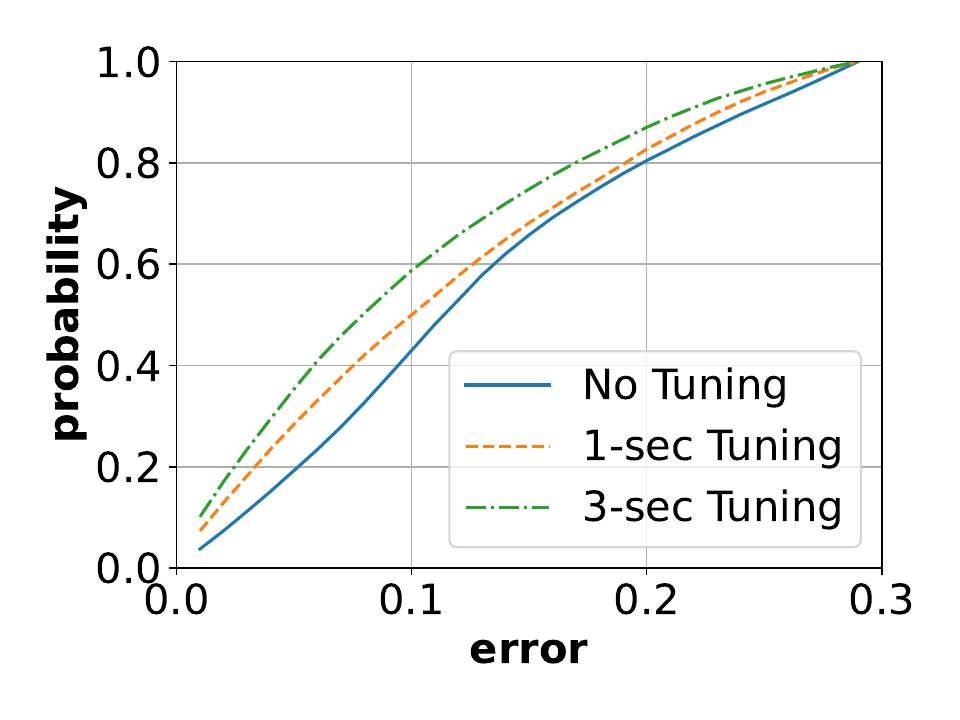} \label{fig:eval1}} \hfill
\subfloat[Accuracy of better mode inference.]{\includegraphics[width=0.49\linewidth]{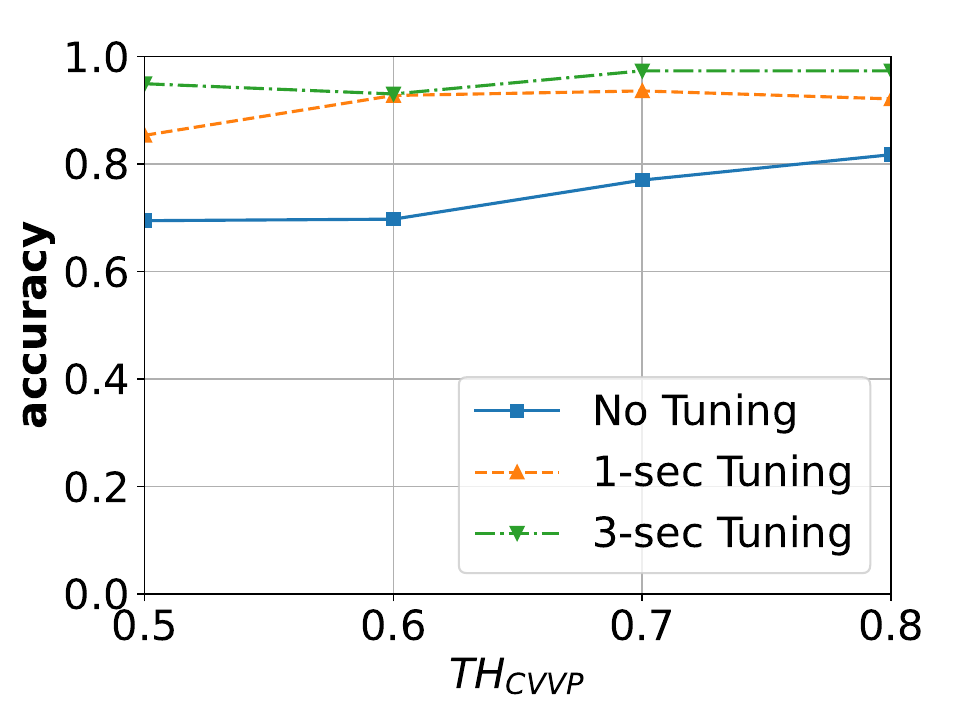} \label{fig:eval2}} \hfill
\caption{Performance of CVVP estimation and mode inference.}
\label{fig:eval}
\end{figure}

Fig.~\ref{fig:eval1} shows the cumulative distribution function (CDF) of the error of predicted $CVVP_i$ for each of the three validation schemes: No Tuning, 1-sec Tuning, and 3-sec Tuning. All of them have an error within 0.15 for about 70\% of the time, and an error within 0.25 for more than 90\% of the time.

Unsurprisingly, the error decreases as more data are used for fine-tuning. Even without any tuning, it still achieves a mean error of 0.19. For 1-sec Tuning and 3-sec Tuning, the mean errors are 0.14 and 0.12, respectively. Considering that the range of $CVVP_i$ is $(0, 1]$, these errors may not be very small. However, it is important to note that $CVVP_i$ will be binarized and stabilized to $\overline{CVVP_t}$ before being used to control view mode switching. For example, a $CVVP_i$ of 0.7 (with its ground truth being 0.9) will be binarized to 1---the same as the binarized ground truth if the threshold $TH_{CVVP}$ (Section~\ref{sec:cvvp_2}) is 0.6. Thus, errors of this level do not prevent 360TripleView from overall accurately inferring the better mode (Section~\ref{sec:eval_53}).

\subsection{Accuracy of Better Mode Inference} \label{sec:eval_53}
Fig.~\ref{fig:eval2} shows the accuracy of 360TripleView in inferring the better mode between \strong ~and \weakman. Even without any tuning, the accuracy is still around 80\%. When 1-sec Tuning is used, the accuracy is raised to above \textbf{90\%} most of the time. The mean accuracy of each scheme is 74\%, 91\% and 96\%, respectively. The threshold $TH_{CVVP}$ varying from 0.5 to 0.8 has no significant impact on the accuracy.

\subsection{Overall Content-Importance}
We compare the overall content-importance when using different \use ~determination strategies: \strong ~ONLY, \weakman ~ONLY, and our 360TripleView. Note that for some videos, 360TripleView infers that \strong ~is the better mode throughout the video, resulting in the same content-importance as the first baseline (\strong ~ONLY). To focus on the performance difference, we exclude those videos and present the average content-importance of the remaining videos whose content-importance varies with \use ~determination strategies.

\textbf{Impact of \use ~Determination Strategies:} The impact of \use ~determination strategies when $TH_{CVVP}=0.6$ is shown in Table~\ref{tab:1}. It demonstrates that our 360TripleView achieves higher overall content-importance than the other \use ~determination strategies (\strong ~ONLY, \weakman ~ONLY) in almost all cases, when the saliency detection strategy (CubePad, ATSal, Pano2Vid) and the tuning time are held constant. Similar results are observed when $TH_{CVVP}=$ 0.5 (Table~\ref{tab:2}) and 0.7 (Table~\ref{tab:3}). 360TripleView's performance may be suboptimal without tuning, and the reason has been clarified in Section~\ref{sec:tuning}.

\begin{table}[h] \centering
\caption{Overall content-importance ($TH_{CVVP} = 0.6$). The best performance value is marked in \textbf{bold}.}
\label{tab:1}
\begin{tabular}{@{}llrrr@{}} \toprule
& & \textbf{CubePad} & \textbf{~~~ATSal} & \textbf{Pano2Vid} \\ \midrule

\multirow{3}{*}{\textbf{No Tuning}} & {\strong} & \textbf{0.293} & 0.411 & 0.782 \\
& {\textsc{Auto}$^{\textsc{opt}}$/\textsc{Man}} & 0.230 & 0.387 & 0.727 \\
& {360TripleView} & 0.115 & \textbf{0.498} & \textbf{0.798} \\ \midrule

\multirow{3}{*}{\textbf{1-sec Tuning}} & {\strong} & 0.089 & 0.187 & 0.594 \\
& {\textsc{Auto}$^{\textsc{opt}}$/\textsc{Man}} & 0.086 & 0.157 & 0.574 \\
& {360TripleView} & \textbf{0.119} & \textbf{0.230} & \textbf{0.723} \\ \midrule

\multirow{3}{*}{\textbf{3-sec Tuning}} & {\strong} & 0.112 & 0.218 & 0.603 \\
& {\textsc{Auto}$^{\textsc{opt}}$/\textsc{Man}} & 0.098 & 0.224 & 0.580 \\
& {360TripleView} & \textbf{0.189} & \textbf{0.239} & \textbf{0.720} \\
\bottomrule
\end{tabular}
\end{table}

\begin{table}[h] \centering
\caption{Overall content-importance ($TH_{CVVP} = 0.5$).}
\label{tab:2}
\begin{tabular}{@{}llrrr@{}} \toprule
& & \textbf{CubePad} & \textbf{~~~ATSal} & \textbf{Pano2Vid} \\ \midrule

\multirow{3}{*}{\textbf{No Tuning}} & {\strong} & 0.150 & 0.208 & 0.582 \\
& {\textsc{Auto}$^{\textsc{opt}}$/\textsc{Man}} & 0.137 & \textbf{0.215} & 0.599 \\
& {360TripleView} & \textbf{0.199} & 0.211 & \textbf{0.636} \\ \midrule

\multirow{3}{*}{\textbf{1-sec Tuning}} & {\strong} & 0.123 & 0.145 & 0.507 \\
& {\textsc{Auto}$^{\textsc{opt}}$/\textsc{Man}} & 0.097 & \textbf{0.160} & 0.482 \\
& {360TripleView} & \textbf{0.188} & 0.155 & \textbf{0.613} \\ \midrule

\multirow{3}{*}{\textbf{3-sec Tuning}} & {\strong} & 0.094 & 0.151 & 0.545 \\
& {\textsc{Auto}$^{\textsc{opt}}$/\textsc{Man}} & 0.068 & 0.130 & 0.512 \\
& {360TripleView} & \textbf{0.154} & \textbf{0.170} & \textbf{0.647} \\
\bottomrule
\end{tabular}
\end{table}

\begin{table}[h] \centering
\caption{Overall content-importance ($TH_{CVVP} = 0.7$).}
\label{tab:3}
\begin{tabular}{@{}llrrr@{}} \toprule
& & \textbf{CubePad} & \textbf{~~~ATSal} & \textbf{Pano2Vid} \\ \midrule

\multirow{3}{*}{\textbf{No Tuning}} & {\strong} & \textbf{0.368} & 0.616 & 0.855 \\
& {\textsc{Auto}$^{\textsc{opt}}$/\textsc{Man}} & 0.133 & \textbf{0.809} & 0.877 \\
& {360TripleView} & 0.056 & 0.716 & \textbf{0.962} \\ \midrule

\multirow{3}{*}{\textbf{1-sec Tuning}} & {\strong} & 0.097 & 0.268 & 0.652 \\
& {\textsc{Auto}$^{\textsc{opt}}$/\textsc{Man}} & 0.092 & 0.246 & 0.640 \\
& {360TripleView} & \textbf{0.144} & \textbf{0.299} & \textbf{0.773} \\ \midrule

\multirow{3}{*}{\textbf{3-sec Tuning}} & {\strong} & 0.141 & 0.296 & 0.657 \\
& {\textsc{Auto}$^{\textsc{opt}}$/\textsc{Man}} & 0.081 & 0.311 & 0.660 \\
& {360TripleView} & \textbf{0.242} & \textbf{0.312} & \textbf{0.781} \\
\bottomrule
\end{tabular}
\end{table}

\textbf{Impact of Saliency Detection Strategies:} 360TripleView focuses on \use ~determination and does not propose a new saliency detection solution. When \use ~becomes \strong ~or \weak, an existing saliency detection approach is employed (Fig.~\ref{fig:overview1}). Comparing CubePad with ATSal, we observe that the latter generally achieves higher content-importance. This is because both approaches infer content-importance based on saliency detection, but ATSal, being a more recent work, combines global and local visual features to predict saliency more accurately compared to previous methods. However, when comparing ATSal with Pano2Vid, a significant difference is still evident. This suggests that the algorithm-generated $(\psi, \theta)$ values are currently quite distinct from the human-labeled ones, indicating room for improvement. While saliency detection falls outside the scope of this paper, we consider it as a potential area for future work.

\textbf{Impact of Tuning:} We observe that when the \use ~determination and saliency detection strategies are held constant, the performance does not always increase with tuning time. This is because the no-tuning model wrongly uses \strong ~consistently on many videos. As previously mentioned, if 360TripleView consistently uses \strong ~on a video, it is essentially equivalent to \strong ~ONLY, so we exclude the video. Consequently, the no-tuning model excludes many videos, which may make the average content-importance of the remaining videos higher.

\section{User Study} \label{sec:user}
\subsection{User Study Settings}
We design and implement an online platform, utilizing it to conduct a user study. A total of 25 participants (21 males and 4 females) are recruited, which is a larger scale compared to many existing works (no user study~\cite{su2016pano2vid,su2017making,yu2018deep,lee2018memory,wang2020attention}, 14 users~\cite{pavel2017shot,cha2020enhanced}, 15 users~\cite{wang2020transitioning360}, 18 users~\cite{lin2017outside,hu2017deep}, 20 users~\cite{kang2019interactive}). The dataset used for evaluation is the Pano2Vid dataset, and we select 6 videos that represent diverse content, including tours (hiking/driving), sports (outdoor/indoor), and parades (daytime/nighttime).

\textbf{User Ratings:} Each participant watches each video under three view modes sequentially: 1) \man ~ONLY, 2) \weakman ~ONLY, and 3) 360TripleView. Participants are asked to rate the content-importance for each video and each mode on a scale from 0 (worst) to 10 (best).

\subsection{User Study Results}
\begin{figure}[h] \centering
\subfloat[Favorite mode.]{\includegraphics[width=0.49\linewidth]{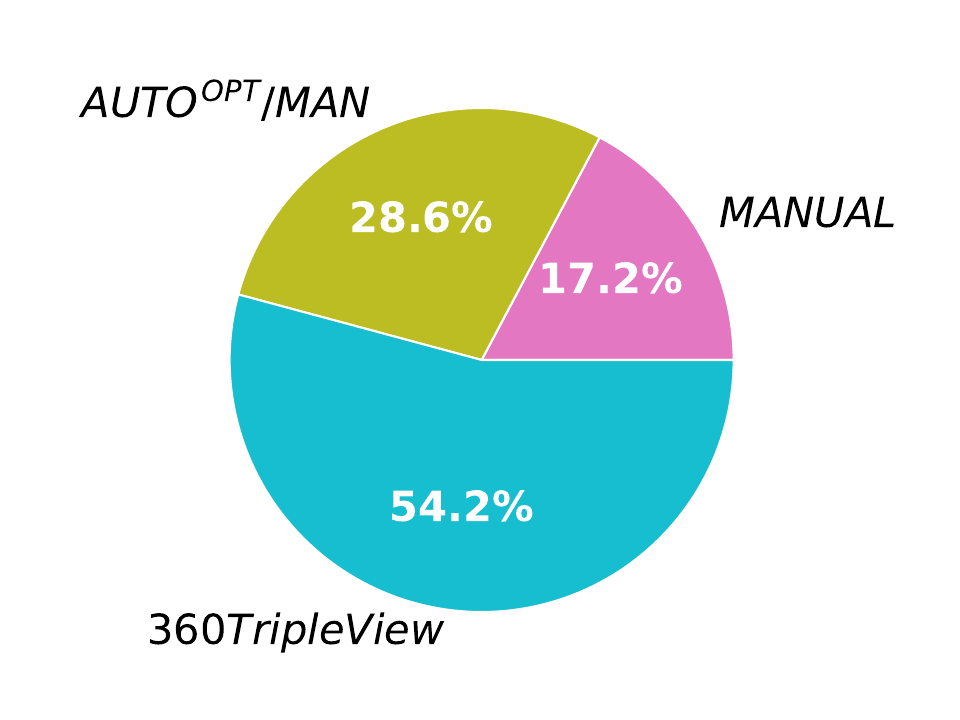} \label{fig:user_1}} \hfill
\subfloat[User rating distribution (green: mean, orange: median).]{\includegraphics[width=0.49\linewidth]{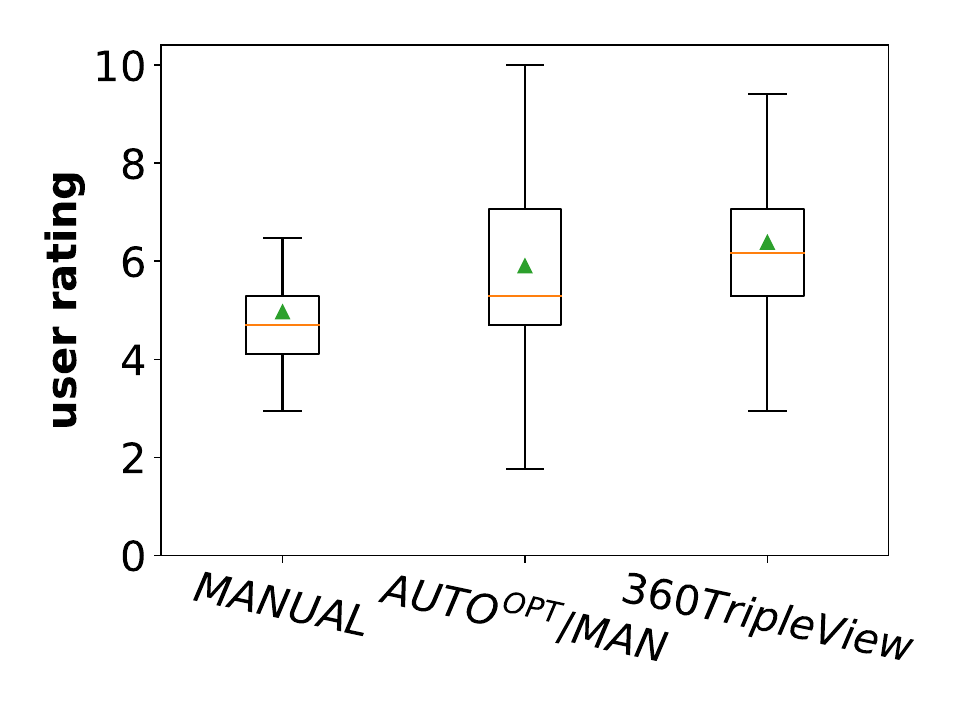} \label{fig:user_2}} \hfill
\caption{User study results.}
\label{fig:user}
\end{figure}

\textbf{Favorite Mode:} If a view mode receives a higher rating than the other two modes from a participant, it is regarded as the participant's favorite mode. Fig.~\ref{fig:user_1} shows the statistics: \man ~ONLY is favored 17.2\% of the time, \weakman ~ONLY is favored 28.6\% of the time, while 360TripleView emerges as the clear winner, being the favorite mode 54.2\% of the time.

\textbf{User Ratings:} The distribution of user ratings for content-importance in each view mode is presented in Fig.~\ref{fig:user_2}. The mean ratings for \man ~ONLY, \weakman ~ONLY and 360TripleView are 4.96, 5.90 and 6.38, respectively. The corresponding median ratings are 4.71, 5.29 and 6.18. It is evident that 360TripleView receives the highest ratings in both measures.

\section{Discussion and Future Work} \label{sec:discussion}
\textbf{Large-Scale Evaluation:} 360TripleView has showcased exceptional results during the evaluation using the Pano2Vid dataset. This dataset comprises 15 videos with 6 $(\psi, \theta)$ labels per frame. It is crucial to note that while other datasets may contain a greater number of videos and annotators, they either utilize 2D videos, or \vid s watched by viewers in a sole \man ~mode, rendering them irrelevant to our study on \vid ~view mode switching (between \strong ~and \weakman). Thus, Pano2Vid is the only dataset that meets our criterion.

Moreover, it is important to emphasize our unwavering commitment to collecting a large \vid ~dataset. However, this task remains ongoing as it necessitates a significantly greater investment of time compared to existing datasets. The process involves engaging multiple participants who meticulously examine every viewing direction of each 360\degree ~view frame by frame, resulting in an arduous and extensive workload. While we have utilized the Pano2Vid dataset for evaluation purposes, it is vital to acknowledge that our own designs and prototypes stand at the vanguard of innovation, and are a profoundly positive and groundbreaking endeavor.

\textbf{CVVP with Motion Considered:} Currently, 360TripleView estimates CVVP for each individual frame based on spatial features only, without considering motion or temporal features across frames. We attempted to incorporate motion, but the unavailability of datasets with ground truth CVVP considering motion hindered our progress. However, it is important to highlight that even when utilizing only spatial features, 360TripleView has demonstrated an accuracy exceeding 90\% in inferring the better mode and has yielded substantially higher content importance in comparison to existing approaches. Furthermore, with the availability of relevant datasets, we can readily enhance 360TripleView by retraining the CVVP regression model using the updated ground truth.

\section{Conclusion} \label{sec:conclusion}
In this paper, we have presented the design, implementation, and evaluation of 360TripleView, a groundbreaking view management system for \vid ~viewing. It offers three view modes and leverages automatic inference to select the better mode between \strong ~and \weakman ~to enhance viewers' overall content-importance. We introduce the concept of CVVP and develop a deep learning-based approach to estimate CVVP automatically and infer the better view mode accurately. Our evaluation results demonstrate that 360TripleView achieves an accuracy above 90\% in inferring the better mode and results in significantly higher content-importance compared to existing approaches.

\bibliographystyle{abbrv-doi}
\bibliography{main}

\end{document}